\PassOptionsToPackage{pdftex,
pdfencoding=auto,
pdfnewwindow=true,
pdfusetitle=true,
bookmarks=true,
bookmarksnumbered=true,
bookmarksopen=true,
pdfpagemode=UseThumbs,
bookmarksopenlevel=1,
pdfpagelabels=true,
breaklinks=true,
colorlinks=true,
}{hyperref}
\documentclass[prx,aps,10pt,superscriptaddress,floatfix,superscriptaddress,notitlepage,nofootinbib,twocolumn]{revtex4-2}
\usepackage[OT1]{fontenc}
\usepackage{nccmath, amssymb, amsthm, mathtools}
\usepackage{bbold}
\usepackage{dsfont}
\usepackage[utf8]{inputenx}
\usepackage[caption=false]{subfig}
\usepackage[export]{adjustbox}
\usepackage{natbib}
\usepackage{verbatim}
\usepackage{soul}
\usepackage{braket}
\usepackage[normalem]{ulem}

\usepackage{graphicx}
\usepackage[dvipsnames]{xcolor}
\definecolor{purple}{RGB}{128,0,128}
\definecolor{ultramarine}{RGB}{63, 0, 255}
\definecolor{medblue}{RGB}{0, 0, 100}
\definecolor{googleblue}{RGB}{34, 0, 204}
\definecolor{panblue}{RGB}{0,24,150}
\definecolor{carmine}{RGB}{150, 0, 24}
\definecolor{gray}{RGB}{150, 150, 150}
\definecolor{darkgreen}{RGB}{0, 80, 0}

\usepackage[]{hyperref}
\hypersetup{linkcolor=carmine,citecolor=darkgreen,urlcolor=googleblue,anchorcolor=OliveGreen}
\usepackage[page,title, titletoc]{appendix}
\usepackage[style=american,autopunct=true]{csquotes} 

\newcommand{\dani}[1]{\textcolor{red}{#1}}
\newcommand{\maria}[1]{{\color{blue} #1}}

\usepackage{newtxtext}
\usepackage{paralist}

\usepackage{microtype}
\microtypecontext{spacing=nonfrench}
\microtypesetup{
expansion={true,nocompatibility},
protrusion={true,nocompatibility},
activate={true,nocompatibility},
tracking=false,
kerning=true,
spacing={true}
}

\usepackage[all=normal,floats=tight,mathspacing=tight,wordspacing=tight,paragraphs=tight,tracking=tight,charwidths=tight,mathdisplays=normal,sections=normal,margins=normal]{savetrees}

\newtheorem{theorem}{Theorem}[section]

\newtheorem{proposition}[theorem]{Proposition}

\begin{document}%
\count\footins = 1000
\title{Certifying Randomness or its Lack Thereof for General Network Scenarios}

\author{Maria Ciudad Alañón}
\affiliation{Perimeter Institute for Theoretical Physics, Waterloo, Ontario, Canada, N2L 2Y5}
\affiliation{Department of Physics and Astronomy, University of Waterloo, Waterloo, Ontario, Canada, N2L 3G1}
\author{Daniel Centeno}
\affiliation{Perimeter Institute for Theoretical Physics, Waterloo, Ontario, Canada, N2L 2Y5}
\affiliation{Department of Physics and Astronomy, University of Waterloo, Waterloo, Ontario, Canada, N2L 3G1}
\author{Andrew Watford}
\affiliation{Department of Physics and Astronomy, University of Waterloo, Waterloo, Ontario, Canada, N2L 3G1}
\author{Elie Wolfe}
\affiliation{Perimeter Institute for Theoretical Physics, Waterloo, Ontario, Canada, N2L 2Y5}
\affiliation{Department of Physics and Astronomy, University of Waterloo, Waterloo, Ontario, Canada, N2L 3G1}

\begin{abstract}
The certification of intrinsic randomness is foundational to quantum information theory and central in many practical applications thereof, such as in the generation of unquestionably random numbers and in cryptographic protocols. Device-independent randomness certification based on violations of Bell inequalities has been thoroughly investigated within the standard Bell scenario. In this work, we aim to extend this line of research by exploring randomness certification in more general causal structures—namely, \emph{network scenarios}. To address this task, we demonstrate how the computational tool known as \textit{the inflation technique} can be adapted. As proof of concept, we use inflation to certify randomness relative to a beyond-quantum adversary for sample probability distributions obtained in the bilocality and triangle scenarios. Complementarily, we also provide computational methods for the problem of certifying an \emph{absence} of randomness, which should not be conflated with certifying the classicality of a given probability distribution. We conclude with a discussion of conceptual subtleties regarding randomness certification in networks, highlighting important open problems in this nascent research field.

\end{abstract}
\maketitle

\section{Introduction}

Intrinsic randomness -- as opposed to randomness due to ignorance -- is a core concept in quantum information theory. Intrinsic randomness is both insightful from the theoretical point of view and a practical resource in numerous applications, such as secure communication and quantum key distribution~\cite{acin2006bell,barrett2005no}. To generate truly random numbers, one requires access to processes that are fundamentally unpredictable~\cite{herrero2017quantum}. Note that, from a foundational perspective, \emph{no} classical systems can exhibit intrinsic randomness, since their dynamics are entirely deterministic. Any apparent randomness in these systems arises only from a lack of knowledge about the underlying description, otherwise known as epistemic randomness. By contrast, quantum theory allows for \emph{intrinsic} randomness~\cite{acin2016certified}.  

The possibility of \textit{true} randomness arising from the nonclassical behaviors predicted by quantum theory was originally proposed in Ref.~\cite{colbeck2009quantum} and proved in Ref.~\cite{pironio2010random}. Such proofs show, in a device-independent manner, how the unpredictability of the outcomes of a quantum correlation (relative to any eavesdropper) follows from the observation of Bell inequality violations. 
The quantitative relationship between nonclassicality and randomness was generalized for different Bell inequalities in Refs.~\cite{acin2012randomness,wooltorton2023expanding}. Randomness has also been studied from the point of view of monogamy of entanglement; Ref.~\cite{augusiak2014elemental} shows that, whenever two parties violate certain Bell inequalities, it follows that there cannot be any third party perfectly correlated with either of them. In addition to theoretical studies, there have also been several experimental implementations~\cite{shalm2021device,li2021experimental,gomez2018experimental}.

Note that here, we are always conceiving randomness as the impossibility that an eavesdropper could predict the outcomes of a measurement. In general, such an adversarial notion of randomness cannot always be established merely from the fact that a correlation is nonclassical. Nonclassicality implies the lack of any (local) hidden variable model. Although the existence of a hidden variable model would imply perfect predictability by an eavesdropper, the converse is not true. This is demonstrated in Refs.~\cite{acin2016necessary,ramanathan2024maximum} where specific Bell inequalities are identified such that these inequalities can violated up to the maximum algebraic non-signalling value all while maintaining perfect predictability by an eavesdropper, hence constituting examples of nonclassicality without randomness. \citet{acin2016necessary} term this phenomenon \enquote{bound randomness}.

More recently, researchers have begun to explore the possibility of randomness certification outside of the standard bipartite \textsf{Bell} scenario. For example, randomness certification has been explored in the tripartite \textsf{Bell} scenario, and the intrinsic randomness in those correlations can be leveraged to enhance the security of device-independent cryptographic protocols~\cite{li2024randomness, grasselli2023boosting, woodhead2018randomness}.
Ref.~\cite{polino2024experimental} shows that the broadcast scenario enhances the robustness of certifiable randomness. Of special interest to us, however, is the emerging line of research regarding randomness certification in more complex scenarios, which have several independent sources, known as quantum networks~\cite{sekatski2023partial, minati2025experimental}. %

In this work, we propose adapting the Inflation Technique~\cite{Wolfe_2019} for the purpose of certifying randomness in networks as a foundational question.\footnote{In particular, here we study scenarios in which the sources are directly connected to the parties (there are no intermediate latents). These are known as exogenous scenarios.} When certifying randomness, it is important to specify the assumptions regarding the adversary that are taken into account when assessing whether a process is predictable or not. As motivated in Ref.~\cite[Sec. II]{pironio2013security}, we consider a potential eavesdropper, Eve, with the ability to ``listen to'' but not ``control'' the sources. Formally, we plausibly imagine that Eve has the ability to measure a share of each source but not to prepare those sources. %
This distinction can be omitted in the case of the standard Bell scenario when considering private settings. The reason is that it is a special case where the joint probability distributions of Alice, Bob and Eve given the settings, i.e. the set of all compatible $P_{A,B,E|X,Y}$, is identical regardless of whether Eve is listening to or controlling the source. However, this is not the case when considering the standard Bell scenario with \emph{public} settings, nor when considering general networks, as explained in Appendix~\ref{app: controlling-listening}.

In our case, we always allow for the settings of every party to be publicly available, such that Eve has access to all of them. The reason for this choice is that settings can always be dilated to bipartite sources. Indeed, most experimental implementations of random settings have devices that physically influence the measurement apparatus while also sending a record of their internal state to a digital recorder. Thus, settings as sources is not merely a mathematical equivalence but perhaps a more accurate causal characterization. Upon appreciating settings as records from bipartite sources, it no longer makes sense to hide them from the eavesdropper. When considering generic networks, why should some sources have security privileges over other sources? See Appendix~\ref{app: controlling-listening} for further discussion.

Another common assumption in randomness certification that we also adopt here is the \textit{closure of laboratories}, i.e., we will always presume that the unseen adversary cannot directly access any information obtained from processes performed privately inside the laboratories.
Recently,~\citet{minati2025experimental} have studied randomness certification in networks in which different subsystems (from different sources) have exclusively one observed party in their causal future. When imagining an unseen eavesdropper,
do we allow the eavesdropper to access those different subsystems after they have interacted, or do we restrict the access of the eavesdropper to the original sources? In Ref.~\cite{minati2025experimental}, they allow for the eavesdropper to access the post-interaction composite system.\footnote{The potential interactions outside the laboratory corresponds to what~\citet{minati2025experimental} call a ``strong eavesdropper''. In terms of the causal formalism, this means changing the causal structure by adding intermediate latent nodes, as considered in Ref.~\cite{centeno2024significance}.} Our perspective, however, is that allowing those interactions prior to eavesdropping conflicts with the assumption of the closure of laboratories since those interactions can be restricted to occur in a single laboratory without loss of generality, and as such we reject the alternative paradigm as an overly strong causal formulation of the potential adversary (see Appendix~\ref{app: strong vs weak} for a more detailed discussion).

There is one final assumption\footnote{Strictly speaking, there is a further assumption in our analysis that cryptographers are sensitive to, namely, that the events are identically and independently distributed, i.e., the IID assumption. The IID assumption negates the possibility of coherent attacks in which Eve and the devices can act differently in each round~\cite{Metger2023}. The IID assumption is baked into the very framework of causal inference that guides all the analysis herein, per positing the existence of a joint distribution of the outcomes of the observed parties alongside the eavesdropper. While randomness certification has been studied in the beyond-IID paradigm in Bell scenarios \cite{van2019correlations,zhang2018certifying}, that has been shown to be impossible in network scenarios~\cite{weilenmann2025memory}.} that must be articulated in order to specify the scope of the eavesdropper's potential to predict the observed outcomes. Namely, one must decide whether the randomness is certified against a quantum or beyond-quantum adversary. Essentially, do we want to assume that quantum theory is true, or do we want to lean into causal paranoia and seek to certify randomness with respect to an adversary limited by \emph{any} future physical theory? Happily, the Inflation Technique is applicable for either paradigm \cite{Wolfe_2019,wolfe2021quantum}. That said, here we elect to showcase the power of inflation by certifying randomness in networks without assuming quantum theory, that is, we certify unpredictability relative to a beyond-quantum adversary.
In particular, we give upper bounds on the guessing probability (a measure of randomness) using \emph{nonfanout} inflation. 

Complementarily, in this work, we also investigate the task of certifying \emph{lack} of randomness for a given party in a network. As mentioned before, the nonclassicality of a probability distribution does not guarantee the presence of randomness (as shown in the standard \textsf{Bell} scenario by Ref.~\cite{ramanathan2024maximum}). In this work, we provide methods to certify that the adversary definitely \emph{can} predict the measurement outcomes of a party in a network while observing a  nonclassical probability distribution. The first method consists of  %
constructing causal models in which the party that is certified to not exhibit randomness receives only classical systems while reproducing the (nonclassical) probability distribution. In the cases of the \textsf{bilocality} and \textsf{triangle} scenarios, this notion coincides with causal modelling availing only one nonclassical source alongside other \emph{classical} source(s). The second method consists of viewing the party that is certified to not exhibit randomness as a player in a Bell scenario (or, in other words, a player who is receiving information from \emph{one} nonclassical source in addition to other classical sources) such that the Bell scenario is embedded within the causal model of the actual causal structure under consideration. By embedding Bell scenarios within network scenarios, we can piggyback on proofs of lack of randomness in Bell scenarios to prove lack of randomness in network scenarios.

\section{Certifying randomness via nonfanout inflation}

\begin{figure}
        \subfloat[\label{fig:intro:bell}]{
                \centering
                \includegraphics[width=3.2cm, valign=b]{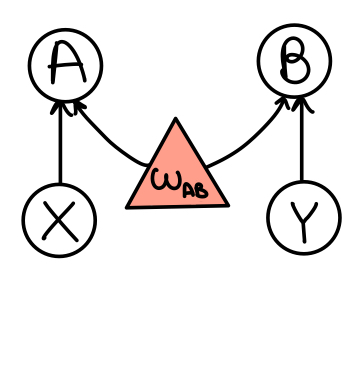}}
        \subfloat[\label{fig:intro:bellEve}]{
                \centering
                \includegraphics[width=3cm, valign=b]{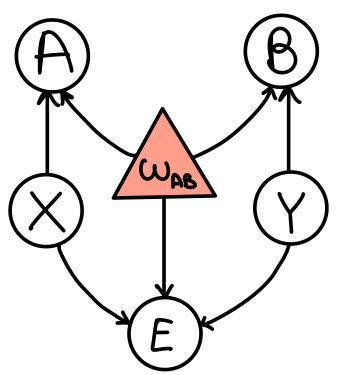}}
        
        \caption{Represenation of the standard \textsf{Bell} scenario (a) and the standard Bell scenario with an eavesdropper, \textsf{Bell+E} (b).}
\label{fig:standard_Bell_Eve} 
\end{figure}

Let us start by stating the problem of randomness certification in the simple case of the standard \textsf{Bell} scenario (see Fig.~\ref{fig:standard_Bell_Eve}a), where two distant parties (Alice and Bob) perform local measurements depending on some settings. In the device-independent paradigm, the unique quantity that is used to certify randomness is the probability distribution over the observed variables (from now on we will use $P_{A,B|X,Y}$ for brevity). To study whether a particular observed correlation $P^{obs}_{A,B|X,Y}$ exhibits intrinsic randomness, we shall consider an adversary, Eve, who has access to the shared resource between Alice and Bob and to their measurement settings, see Fig.~\ref{fig:standard_Bell_Eve}b). Now, the scenario is described by the joint probability distribution of Alice, Bob and Eve given the settings: $P_{A,B,E|X,Y}$. We will say that a correlation shows intrinsic randomness in Alice's outcomes for a given setting when Eve fails to perfectly guess them. This can be formulated as an optimization problem over the set of correlations compatible with the causal structure including Eve, $\mathcal{S}_E$\footnote{We use the same notation for the causal structure and the set of probability distributions compatible with such causal structure.}, that maximizes the guessing probability of Eve, while maintaining the marginal on Alice and Bob to match the observed correlation. That is,

\begin{subequations}
    \begin{align}
        p_{\text{guess}}^{A_x}\coloneqq \; &\max\; \sum_a P_{A,E|X}(a,a|x)\\
        &\mbox{s. t.} \quad P_{A,B,E|X,Y} \in \mathcal{S}_E\\
        &\mbox{and} \quad P_{A,B|X,Y} = P_{A,B|X,Y}^{obs}\,.
    \end{align}
    \label{eq: randomness}

We can also define 
\begin{align}
p_{\text{worst\_guess}}^A\coloneqq \min_x\; p_{\text{guess}}^{A_x}.
\end{align}
\end{subequations}

Thus, we say that there \emph{is} intrinsic randomness in Alice's outcomes if and only if $p_{\text{worst\_guess}}^A\lneq 1$. %
Note that in this formulation we do not specify the mathematical construction of the tested probability distribution $P_{A,B|X,Y}^{obs}$\footnote{Typically, it is assumed that the correlations are quantum, i.e., they can be predicted by the Born rule. In this work, however, we allow for any possible nonclassical probability distribution.} nor the nature of the adversary, quantum or beyond-quantum.\footnote{This assumption will later lead to a different set of probability distributions compatible with the causal structure and therefore different sets over which we maximize.} Furthermore, there are two ways of defining the guessing probability, depending on which settings the eavesdropper is attempting to predict: the average guessing probability, where she tries to guess \emph{every} input, and the fixed-setting guessing probability, where she attempts to guess \emph{any} single input. Note that these two definitions need not coincide. 
Since fixed-setting unpredictability implies average unpredictability, it is a stronger certification, and it is the one we adopt here. Note that the focus of this work is not on \emph{quantifying} randomness, but rather on determining its \emph{presence or absence}.

The mathematical problem defined by Eq.~\ref{eq: randomness} can be generalized to certify randomness in network scenarios. Consider a given network described by a directed acyclic graph (DAG) $\mathcal{G}$. Consider a particular probability distribution $P_{\bar{A}|\bar{X}}^{obs}\in \mathcal{G}$, where $\bar{A}$ is the list of outcomes of the parties and $\bar{X}$ the corresponding settings. Analogously, the optimization problem that we solve to determine whether the correlation has intrinsic randomness or not is:
\begin{subequations}
\label{optimization_networks}
    \begin{align}
        p_{\text{guess}}^{\bar{A}_{\bar{x}}}\coloneqq \; &\max\;\sum_{\bar{a}} P_{\bar{A},E|\bar{X}}(\bar{a},\bar{a}|\bar{x})\\
        &\mbox{s. t.} \quad P_{\bar{A},E|\bar{X}} \in \mathcal{G}_E\\
        &\mbox{and} \quad P_{\bar{A}|\bar{X}} = P_{\bar{A}|\bar{X}}^{obs},
    \end{align}
\end{subequations}
where $\mathcal{G}_E$ denotes set of probability distributions compatible with the DAG including the eavesdropper, $E$, with access to all the sources and all the inputs. Analogously, we say that there \emph{is} inherent randomness in the parties $\bar{A}$ if and only if $p_{\text{worst\_guess}}^{\bar{A}}\lneq 1$. Consequently, this framework can be adapted to certify randomness in any subset of parties although here we focus on certifying single-partite randomness. The most challenging part of this optimization problem is to determine over which set of probability distributions one should optimize, i.e., the set of compatible distributions with $\mathcal{G}_E$. 

In general, the set of correlations compatible with a given causal structure involving multiple independent sources is known to be non-convex, in contrast to the case of the standard \textsf{Bell} scenario (where there is only one source). Thus, the problem of assessing whether there is a distinction between the distributions generated by classical versus nonclassical resources is also more difficult than in the standard \textsf{Bell} scenario. This problem has resulted in the development of computational methods to establish bounds on the set of classical, quantum and post-quantum correlations in networks~\cite{pozas2019bounding, kela2019semidefinite, weilenmann2018non}. The Inflation Technique \cite{Wolfe_2019,wolfe2021quantum} is one of those methods and the one we propose to use to tackle the problem of randomness certification in networks. This approach enables the formulation of optimization problems over outer approximations of the set of compatible distributions. As a result, upper bounds can be derived for the adversary's guessing probability, under the assumption that the marginal distribution observed by the honest parties is fixed. In other words, lower bounds on the amount of certifiable randomness can be obtained. The type of inflation used depends on the nature of the underlying resources: fan-out inflation for classical, quantum inflation for quantum, and nonfanout inflation for post-quantum. This problem is implemented via linear programming for classical and beyond-quantum scenarios, and semidefinite programming for the quantum case. Throughout this manuscript, we assume Eve to be a beyond-quantum adversary, thereby allowing her to be as powerful as possible. Consequently, we consider only nonfanout inflations.

The results presented in this paper using the Inflation Technique were obtained with the package available in Ref.~\cite{boghiu2023inflation}. Given a DAG, the inflation level, and the objective function, this package systematically explores all possible inflations to optimize the objective function. See Appendix~\ref{app: bilocality_inflation} for a detailed explanation of how to implement the first two levels of inflation to certify randomness in the \textsf{bilocality} scenario.

\begin{table}[b]
\begin{tabular}{|l|l|l|l|l|}
\hline
\textbf{Correlation}  & \textbf{Network} & \textbf{Party} & \multicolumn{1}{c|}{\textbf{\begin{tabular}[c]{@{}c@{}}Inflation\\ level\end{tabular}}} & \multicolumn{1}{c|}{\textbf{\begin{tabular}[c]{@{}c@{}}\boldmath{$p_{\text{worst\_guess}}$} \\\textbf{bound}\end{tabular}}} \\ \hline
Fritz-inspired & \textsf{bilocality} & A, B & (2, 2)  & 0.7929 \\ \hline
Entanglement-swapping & \textsf{bilocality} & B  & (2, 3) & 0.9815 
\\ \hline
Entanglement-swapping & \textsf{bilocality} & A, C & (2, 2) & 0.8964
\\ \hline
Fritz's \textsf{triangle} & \textsf{triangle} & A, B & (1, 2, 2) & 0.9879       \\ \hline
Post-quantum & \textsf{triangle} & A, B, C & (2, 2, 3) & 0.8369      \\ \hline
\end{tabular}
\caption{Upper bounds on the fixed-setting worst-case guessing probability for different probability distributions (details of the probability distributions on Appendix~\ref{app: correlations}.) 
}
\label{table: results}
\end{table}

To demonstrate the efficacy of this technique, we utilize inflation to certify the presence of randomness across a range of probability distributions (see Appendix~\ref{app: correlations} for detailed descriptions of the distributions) in the \textsf{bilocality} and \textsf{triangle} networks, with the results of the upper bounds on guessing probabilities for different parties summarized in Table~\ref{table: results}. As a first example, inflation level $(2,2)$ can be used to certify single-partite randomness for Alice and Bob in the \textsf{bilocality} scenario for the distribution inspired by Fritz~\cite{Fritz_2012}, where Charlie’s output matches Bob’s input and Alice and Bob violate CHSH. This distribution illustrates the crucial role of the assumed causal structure in randomness certification: if we were to assume the DAG of a standard three-party \textsf{Bell} scenario with a single source, no randomness could be extracted from this distribution; however, under the assumption of source independence, randomness can be certified.\footnote{Importantly, this remark is closely tight to the assumption of a ``listening'' adversary rather than a ``controlling'' one. See appendix\ref{app: controlling-listening} for a discussion on this} Moreover, in the entanglement-swapping protocol, randomness can also be certified for the middle party, showing that randomness can be certified for settingless parties in network scenarios---a phenomenon that does not occur in the standard \textsf{Bell} scenario. Finally, we also provide examples of randomness certification in the \textsf{triangle} scenario without settings, noting that while one such distribution is compatible with quantum theory, the other is not.

\section{Certifying the lack of randomness via inner aproximation}
\label{sec: lack}

Certifying the absence of randomness for a given party implies predictability of all of that party's measurement outcomes regardless of the measurement setting. Following the formulation in Eq.~\eqref{eq: randomness}, certifying lack of randomness in Alice's measurement implies obtaining $p_{\text{worst\_guess}}^{A}=1$. %
While the Inflation Technique is a powerful tool for certifying the \emph{presence} of randomness in any network, it is not well suited for certifying the \emph{absence} of randomness. A given nonfanout inflation can be used to witness the causal \emph{incompatibility} of the original multipartite correlation upon extending it to include an eavesdropper with $p_{\text{worst\_guess}}^{A} = 1$. However, even if one or more nonfanout inflations are consistent with such a perfect-prediction extension of the correlation, that does not guarantee the existence of an explicit causal model allowing for perfect prediction, even upon considering beyond-quantum causal resources such as those compatible with generalized probabilistic theories~\cite{muller,plavala2023general,PhysRevA.75.032304}.
To certify a \emph{lack} of randomness, therefore, we turn to inner constructions. 

\subsection{Lack of randomness proofs with classical parents}
To certify the absence of single-party randomness we primarily leverage the principle that classical systems cannot exhibit intrinsic randomness. This is formalized as follows:
\begin{proposition}
\label{propositionlack}
    Consider a correlation $P_{\bar{A}|\bar{X}}$ compatible with a given DAG $\mathcal{G}$. If there exists a causal model for $\mathcal{G}$ that reproduces this correlation and in which the party $A_i$ receives only classical sources, then $A_i$ contains no randomness (i.e., an eavesdropper $E$ with access to the sources received by $A_i$ can always predict its outcomes).

\end{proposition}
In practice, to certify the lack of intrinsic randomness in a single party ($A_i$) with respect to a beyond-quantum adversary, our algorithm works as follows: Take as input the probability distribution under consideration, $P_{\bar{A}|\bar{X}}$, and attempt to construct a concrete model wherein all the sources which are causal parents of $A_i$ are classical (while the rest of sources are allowed to be post-quantum resources), such that the causal model ultimately yields the given distribution.

It is important to recognize that the question of whether or not such a causal model can be found is not the same as asking whether or not there is a hidden variable model to obtain the correlation, i.e., to assess whether the probability distribution is classical or not. Certifying the classicality of the correlation is equivalent to finding a causal model in which \emph{all} the sources in the network are classical. Although that would indeed certify the lack of randomness in all the parties, our main point here is that the construction of mixed source type causal models allows us to certify the lack of randomness for individual parties even when the observed correlation is nonclassical.

\subsubsection{Bilocality}
\label{bilocality}
\begin{figure}[t]
\centering
\includegraphics[width=5cm]{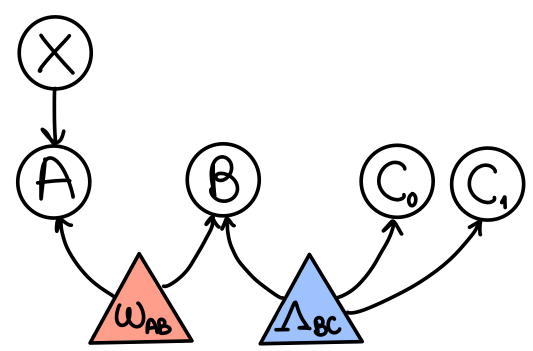}
\caption{Unpacked \textsf{bilocality} scenario for Charlie's settings. The blue triangles represent the classical sources, while the orange ones represent nonclassical sources.}
\label{fig:bilocal_unpacking}
\end{figure}

For the case of the \textsf{bilocality} scenario, ~\citet{ciudad2024escaping} explicitly present a linear program that can be utilised to ascertain the existence of a causal model with one classical source and one nonclassical source. In summary, a correlation $P_{A,B,C|X,Z}$ is compatible with one nonclassical source between $A$ and $B$ and a classical one between $B$ and $C$ in the \textsf{bilocality} scenario if and only if there exists a probability distribution, $Q_{A,B,C_0,C_1|X}$, over the unpacked DAG,\footnote{The notion of unpacking is related to classical explainability, as one imposes the possibility of performing all the measurements of a party simultaneously~\cite{Fine1982}.} represented in Fig.~\ref{fig:bilocal_unpacking}, which satisfies the no-signalling and independence constraints coming from the causal structure and the compatibility constraint relative to the \textsf{bilocality} scenario. Therefore, we can ensure that there is \emph{no randomness for Charlie} from $P_{A,B,C|X,Z}$ in the \textsf{bilocality} scenario if:

\begin{subequations}\label{LP_biloc}
\begin{align}\nonumber
\exists\;\; &Q_{A,B,C_0,C_1|X}\geq 0\quad\textrm{ s.t.}\\
\label{LP_biloc_NSX}&Q_{B,C_0,C_1|X}=Q_{B,C_0,C_1}\\
\label{LP_biloc_fac}&Q_{A,C_0,C_1|X} = Q_{A|X}Q_{C_0,C_1}\\
\label{LP_biloc_rec}&Q_{A,B,C_z|X}(a,b,c|x)= P_{A,B,C|X,Z}(a,b,c|x,z)
\end{align}
\end{subequations}

Using this optimization, it can be proven, for instance, that the previously mentioned example in the \textsf{bilocality} scenario inspired by Fritz does not have any randomness for Charlie, as the protocol to achieve that correlation can be explained with one classical source between Bob and Charlie. Additionally, Ref.~\cite{ciudad2024escaping} defined a notion of genuine network correlations named Minimal Network Nonlocality (MNN), which includes the correlations that ``cannot be modeled by allowing
all the sources in the network to be classical, while it is compatible with \emph{all} causal interpretations wherein exactly one of the sources is a beyond-quantum resource and the others are classical''. Therefore, all the MNN correlations provided there are also inside the set of interesting cases where we can certify the single-party lack of randomness for both extreme parties of the \textsf{bilocality} scenario, i.e., Alice and Charlie, while being nonclassical correlations.

\subsubsection{Triangle}
\label{triangle}

\begin{figure}
       \subfloat[\label{fig:tribil:tri}]{
                \includegraphics[width=3.7cm]{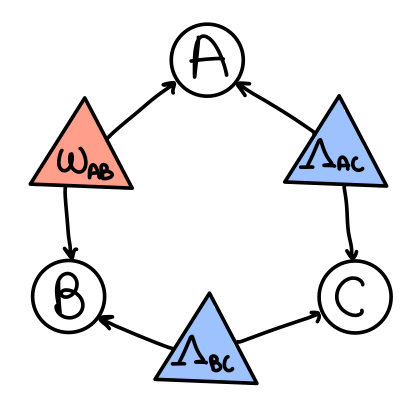}}
        \subfloat[\label{fig:tribil:bil}]{
                \includegraphics[width=4.1cm]{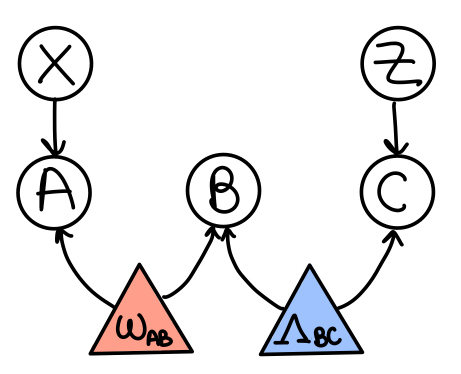}}
        \caption{\textsf{Triangle} scenario with two classical sources and \textsf{bilocality} scenario with one classical source.}
\label{fig:triangle-bilocal} 
\end{figure}

To certify the absence of randomness for a single party in the \textsf{triangle} scenario, following Proposition~\ref{propositionlack}, we construct causal models with two classical sources and one beyond-quantum nonclassical source. Our approach builds on top of the linear program developed for the \textsf{bilocality} scenario, Eqs.~\eqref{LP_biloc}. The key reason is that a correlation $P_{A,B,C}$ compatible with one nonclassical and two classical sources in the \textsf{triangle} network can be understood as a post-selection of a correlation compatible with the \textsf{bilocality} scenario where one of the sources is classical and the other, nonclassical, but where the setting values remain unobserved. In particular, that post-selection is the one where the inputs of the extreme parties of the hypothetical \textsf{bilocality} scenario are constrained to be identical because the \textsf{triangle} features a common source between those parties. Moreover, the settings of the extreme parties of the hypothetical \textsf{bilocality} scenario are not only post-selected to coincide but are also unobserved, thereby acting as a classical latent source.

Formally, this situation is captured by a bilinear program in which the cardinality of one of the two classical sources (specifically, the source connecting the extreme parties in the \textsf{bilocality} scenario) is treated as an explicit and adjustable parameter that determines the model search space. For certain such cardinality choice, the resulting bilinear program can be efficiently solved using Gurobi~\cite{gurobi}. Let us present the mathematical formulation of the bilinear program that implies the certification of lack of randomness for Charlie from a given correlation $P_{A,B,C}$. To do so, we need to consider the \textsf{triangle} and the \textsf{bilocality} networks both with the only nonclassical source between Alice and Bob (see Fig.~\ref{fig:tribil:tri} and \ref{fig:tribil:bil}, respectively). Thus, it can be formulated as follows; there is \emph{no randomness for Charlie} from $P_{A,B,C}$ in the \textsf{triangle} scenario if relative to some hidden cardinality $|Z|=|X|=d$,

\begin{subequations}\label{LP_triangle}
\begin{align}\nonumber
\exists\;\; &\{ Q_{A,B,C_1,C_2,...,C_{d}|X}\geq 0,\quad Q'_{Z}\geq 0\} \quad\textrm{ s.t.}\\ 
\label{triangle_NSX}&Q_{B,C_1,C_2,...,C_{d}|X}=Q_{B,C_1,C_2,...,C_{d}}\\
\label{triangle_fac}&Q_{A,C_1,C_2,...,C_{d}|X}=Q_{A|X}Q_{C_1,C_2,...,C_{d}}\\
\label{triangle_rec}&\sum_{z{=}1}^d Q'_{Z}(z)Q_{A,B,C_{z}|X}(a,b,c|x{=}z)=P_{A,B,C}(a,b,c)
\end{align}
\end{subequations}

Of course, if the bilinear program fails to find such a model, one can try again with higher specified hidden cardinality $d$. Notice that, if $P_{A,B,C}$ is asymmetric with respect to the exchange of $B$ and $C$, to certify lack of randomness for Charlie one might want to consider a flipped version of this bilinear program, wherein the Bob-Charlie classical source would be the one to have fixed cardinality instead. To investigate the absence of randomness for parties other than Charlie, the bilinear program can be straightforwardly adapted by placing the nonclassical source between $B$ and $C$ or between $A$ and $C$.

We utilize the formulation as presented in Eqs.~\eqref{LP_triangle} to find a model for the RGB3 distribution proposed in~\cite{boreiri2023towards} that is defined by two parameters ($u$ and $\lambda_0$). To assess the nonclassicality of such distribution, we use the witness proposed in~\cite[Eq. (C4)]{boreiri2023towards}. The distribution with the biggest violation (assuming cardinality of $\Lambda_{AC}$ equal to 2) of that witness that we can certify lack of randomness of corresponds to $u=0.93$ and $\lambda_0=0.693$. Note that increasing the cardinality of $\Lambda_{AC}$ might allow to find a model for a distribution with a bigger violation. Further details of the model can be found on {\texttt{GitHub:mciudada/Randomness}} \cite{repo}.
This proves that there is no randomness in $C$ within RGB3 relative to a beyond-quantum eavesdropper. Furthermore, as RGB3 is a symmetric distribution, analogous models showing lack of randomness in either $A$ or $B$ instead can also be found by relabelling the components of the causal model certifying no randomness in $C$. That is, we find that there is no single-partite randomness for \emph{any} party from the RGB3 distribution given those parameters. 

\subsection{Lack of randomness proofs with a nonclassical parent}
The search of causal models such that all sources pointing into a particular party are classical can be an effective tool for certifying lack of randomness for that party. However, one must appreciate that there exist correlations which lack randomness for a given party such that the lack of randomness cannot be demonstrated by such a construction. Indeed, one need look no further than the \textsf{Bell} scenario. As shown by~\citet{ramanathan2024maximum}, in the \textsf{Bell} scenario there exist correlations which are nonclassical and yet which also lack randomness. As the \textsf{Bell} scenario is comprised of a single source, that means that the outcomes of Alice can be shown to lack randomness despite the impossibility of a causal explanation in which Alice is connected exclusively to classical sources.

In networks, we can similarly construct causal models in which a party is connected to \emph{one} nonclassical source (and, possibly, some other classical sources) in such a manner as to lack randomness. A causal model with classical sources and one nonclassical source can always be thought of as an embedding of a \textsf{Bell} scenario (in a network), where some of the settings may come from the classical hidden sources. %
Then, the idea to prove lack of randomness in the correlation obtained in the network is to construct such a causal model where the embedded Bell scenario is shown to lack randomness.
See Appendix~\ref{app: lack_of_randomness_embeddings} for a mathematical formulation of this feasibility problem for the bilocality and the triangle scenarios.

\section{Discussion}

In this work, we address the foundational problem of certifying the presence or absence of intrinsic randomness in a probability distribution given a causal structure. This problem has been well studied for the case of the standard Bell scenario and here, we transition to more complex causal structures in which more than one independent source are present, i.e., networks. %

For the matter of randomness certification, we propose the use of the Inflation Technique and we show the efficacy of this method by proving randomness in different well-known probability distributions produced in the \textsf{bilocality} and the \textsf{triangle} scenarios, assuming a beyond-quantum adversary. 

On the other hand, for certifying the absence of randomness in networks, we first provide a computational approach that certifies lack of randomness based on the premise that classical systems do not show unpredictability. In particular, we provide causal model constructions for the \textsf{bilocality} and the \textsf{triangle} scenario wherein one of the parties receives exclusively classical systems. This allows us to show that the RGB3 distribution~\cite{boreiri2023towards} (for a particular range of parameters) which is nonclassical relative to the \textsf{triangle} scenario nevertheless lacks single-partite randomness against a beyond-quantum adversary. Secondly -- as shown by~\citet{ramanathan2024maximum} in the standard \textsf{Bell} scenario -- we recognize the possibility that a distribution may (similarly) lack randomness for some party despite resisting explanation in terms of a causal model utilizing only classical sources for that party. Thus, we also provide a method capable of certifying lack of randomness even for a party who must be connected to a nonclassical source to explain the observed correlation. We have provided explicit formulations of this approach in both the \textsf{bilocality} and \textsf{triangle} scenarios. This latter computational method warrants consideration of two subtleties, which we subsequently elaborate: firstly, that single-partite randomness (or lack thereof) is distinct from multipartite randomness (or lack thereof). Secondly, the inapplicability of our techniques for certifying lack of randomness given a correlation wherein the party in question cannot be modeled with all-but-one of their sources taken to be classical.

\emph{Single-party versus multipartite randomness}. Note that the absence of single-partite randomness does not imply the absence of randomness in the joint probability distribution. Indeed, even in Bell scenarios one can find correlations that lack randomness in the outcomes of any individual party but where the joint outcomes of multiple parties are certifiably unpredictable by an eavesdropper. An example of this phenomenon can be encountered when considering the set of correlations which violate the $I_{3322}$ inequality~\cite{sliwa2003symmetries,collins2004relevant}, in particular, by considering the variant of $I_{3322}$ which is symmetric under exchange of parties~\cite[Eq.~(4)]{brunner2008partial}.  $I_{3322}$ \emph{can} be violated by correlations wherein either the outcomes of Alice or Bob are predictable (for all settings) relative to a nonsignalling eavesdropper, but $I_{3322}$ cannot be violated by any correlation such that the \emph{joint} outcomes of Alice and Bob are similarly predictable. With this distinction in mind, note that the programs to witness lack of randomness constructing models where all but one source are classical (as described in Appendix~\ref{app: lack_of_randomness_embeddings} for the \textsf{bilocality} and \textsf{triangle} scenarios) can certify lack of randomness for more than one party at once, whereas the programs constructing models where all the sources received by a party are classical (as described in Sec.~\ref{bilocality} for the \textsf{bilocality} and in Sec.~\ref{triangle} for the \textsf{triangle}) only certify single-partite absence of randomness.

\emph{Missing techniques to certify lack of randomness}. As mentioned, we have introduced satisfiability problems which, when feasible, amount to certifying the lack of randomness for some party. All these satisfiability problems share a deficiency, however. Namely, they only work to certify lack of randomness in some party's outcomes if the observed distribution can be causally modelled while restricting the party in question to have \emph{at most one} nonclassical source among their causal parents. Consider a distribution which resists any such causal explanation. All \emph{Fully Network Nonlocal} correlations in the sense of Ref.~\cite{Pozas_Kerstjens_2022} are of this sort, at least for parties with more than one latent source among their parents (thereby excluding the \textsf{Bell} scenario). Do such correlations necessarily give rise to certifiable randomness? It seems \emph{a priori} plausible that there could exist distributions which lack single-party randomness despite resisting an explanation in terms of a causal model wherein that party only receives at most one nonclassical system. The existence of such distributions would imply the inadequacy of the techniques presented in this work, as the techniques here would be incapable of witnessing that lack of randomness. Could it be that no such distributions exist? If so, how could such a claim be proven?

In this work, we have restricted ourselves to causal structures where the latent nodes do not have parents, i.e., exogenous causal structures. However, it was shown in Ref.~\cite{centeno2024significance} that considering non-exogenous scenarios can make a difference when nonclassical sources are present. Therefore, we leave for future work the study of certifying the presence or absence of randomness in those causal structures that are non-exogenous \emph{before} adding the eavesdropper.

While this work addresses the question of certifying the presence or absence of randomness, the question of \emph{quantifying} the degree of randomness when randomness is present is also important, and should be addressed in future work. Moreover, this work investigates randomness certification in networks purely from a foundational perspective, leaving as an open question whether this sort of randomness has any applications.

\section{Acknowledgements}
\noindent We acknowledge helpful discussions with Yeong-Cherng Liang, Ernest Tan, Marina Maciel Ansanelli, Soham Bhattacharyya and Sadra Boreiri. Research at Perimeter Institute is supported in part by the Government of Canada through the Department of Innovation, Science and Economic Development and by the Province of Ontario through the Ministry of Colleges and Universities. M.C.A. and D.C. acknowledge support from NSERC grants 50505-11449 and 50505-11450.
\vspace{-0.05in}
\setlength{\bibsep}{.2\baselineskip plus .05\baselineskip minus .1\baselineskip}
\bibliographystyle{apsrev4-2-wolfe}
\nocite{apsrev42Control}
\bibliography{refs.bib}
\appendix
\section{``Listening'' versus ``controlling'' eavesdropper}
\label{app: controlling-listening}

\begin{figure}[h]
        \subfloat[\label{fig: listen}]{
                \centering
                \includegraphics[width=3.6cm]{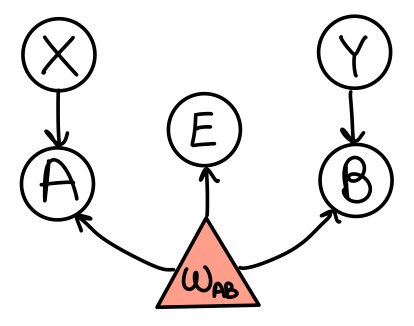}}
        \subfloat[\label{fig: control}]{
                \centering
                \includegraphics[width=3.6cm]{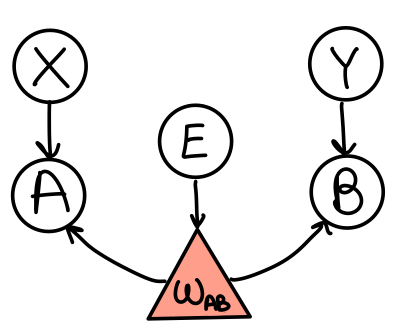}}
        \caption{Different extensions of the standard \textsf{Bell} scenario to include an eavesdropper. In (a) Eve is ``listening to'' the source, whereas in (b) she is ``controlling" the source.}
\label{fig:Bell-Eve} 
\end{figure}

The device-independent paradigm allows one to establish information-theoretic secrecy from experimental correlations without relying on -- or even \emph{verifying} -- the honesty of the supplier of the nonclassical devices. There are different forms of malicious behavior by the device supplier; the different cryptographic attack paradigms correspond to distinct placements of the adversary in the causal structure capturing the minimal security assumptions. In one paradigm, we assume that the source is not being manipulated during the experiment, but we allow for a leaky source in the sense that the internal state may include a system that an eavesdropper can probe and measure. The corresponding causal structure places Eve as a causal descendant of the source. In another paradigm, we imagine that the adversary may actively be tampering and adjusting the source during the experiment, in which case we place Eve as the causal ancestor of the source. These paradigms are illustrated in Fig.~\ref{fig:Bell-Eve}. We refer to these two different models as a ``listening'' or ``controlling'' eavesdropper, respectively. 

\begin{figure}[h]
\centering
\includegraphics[width=5cm]{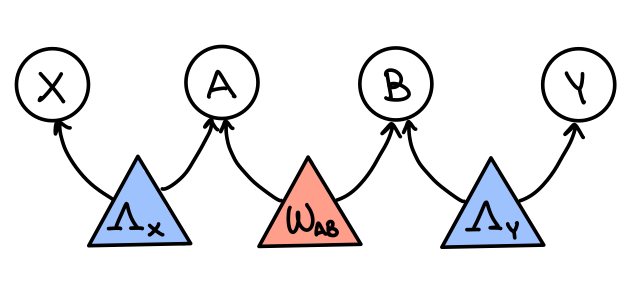}
\caption{Representation of the standard Bell scenario where the settings are produced by classical latent sources.}
\label{fig:Bell-network}
\end{figure}

It is important to note that, in the standard \textsf{Bell} scenario, the two causal structures are observationally equivalent, in the sense that they generate the same set of compatible correlations.  Hence, the results of randomness certification and the quantitative security of cryptographic protocols are invariant regardless of which paradigm one assumes. However, this oft-relied-upon equivalence is predicated on an implicit caveat which we must call attention to, namely, the assumption of private settings. Although this assumption seems reasonable at first glance, it may become questionable when we think about experimental implementations, especially when we think about what it means to generalize the assumptions to more complex network scenarios. Most Bell nonlocality experiments are performed using classical sources of randomness to toggle the settings. Some use quantum random number generators. Regardless, upon recognizing that the setting is manipulated by an external source of some kind which leaves a record for the experimenter, it becomes natural to recast the standard Bell scenario as a three-source network (see Fig.~\ref{fig:Bell-network}). From that perspective, it is unclear why the latent sources producing the settings should be ``privileged'', i.e., specially exempt from Eve's influence. 

If one adopts the paranoid attitude that Eve \emph{controls} all latent sources in the scenario, then randomness certification becomes impossible. For example, in the causal scenario representing experimental implementations of the standard Bell scenario, if Eve controls all three latent sources, she can readily fine-tune them so that Alice and Bob still \emph{observe} a nonsignalling distribution all while she predicts their outputs perfectly. An easy way to appreciate Eve's power to select the outcomes in advance is to realize that she can effectively act as an arbitrary four-way common cause with deterministic causal dependence and Alice and Bob would be none the wiser. Indeed, whatever justifies the experimenters' confidence that the latent sources are causally independent is the very same justification against that possibility of an all-controlling adversary. %

By contrast, if Eve is only ``listening'' to the sources, then randomness can still be certified. This was already emphasized by \citet{pironio2013security}, who pointed out that treating settings as public and adopting the listening model still enables randomness certification. Only the ``listening" paradigm makes sense when we have multiple latent sources, none of which are privileged in terms of being shielded from the adversary. As such, we conclude that the listening model provides the most consistent and conceptually natural adversarial assumption for studying randomness in networks. 
Of course, one may prefer the more paranoid stance that Eve \emph{controls} everything, but in that case we think that one should then forego the possibility of randomness certification even in the standard Bell scenario.

\section{Strong vs weak eavesdropper}
\label{app: strong vs weak}
\begin{figure}[h]
    \subfloat[\label{fig:weakstrong:weak}]{
                \includegraphics[width=3.9cm]{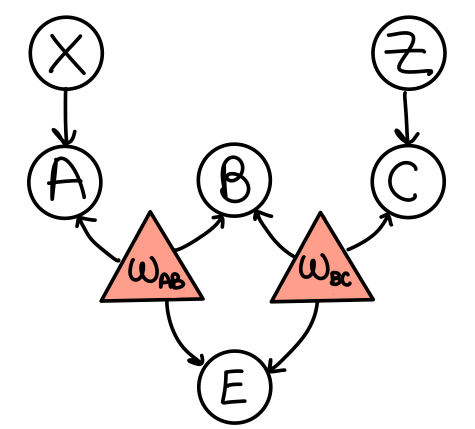}}
        \subfloat[\label{fig:weakstrong:strong}]{
                \includegraphics[width=4cm]{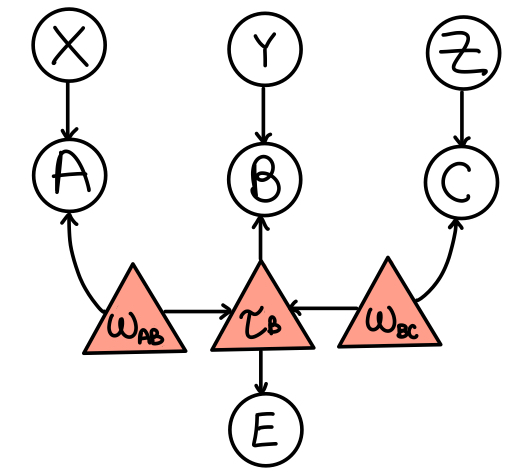}}
        \caption{The \textsf{bilocality} scenario as considered by \citet{minati2025experimental}, namely, within the paradigm wherein settings --- distinct from sources --- have \emph{privileged security}, such that the values of the settings are never learned by any form of adversary. (a) depicts the scenario relative to a weak eavesdropper. (b) depicts the scenario relative to a strong eavesdropper.}
\label{fig:weakstrong} 
\end{figure}

One key assumption to certify randomness or the security of cryptographic protocols in a device-independent manner is \emph{closure of laboratories}. This assumption means that the eavesdropper cannot observe or obtain any information about the processes carried out inside the laboratories of the different honest parties. Recently, the field of randomness certification has considered more general causal structures than the usually studied standard \textsf{Bell} scenario. \citet{minati2025experimental} considered the \textsf{bilocality} causal structure and proposed different eavesdropper models. In particular, they proposed two models named ``weak'' or ``strong'' eavesdropping. The former prohibits interactions of the different subsystems going to the middle party prior to entering said party's laboratory, whereas the latter allows them. In terms of causal structures, the difference amounts to considering an exogenized or non-exogenized scenario when including the eavesdropper; i.e., considering a causal structure without (for the weak) or with (for the strong) intermediate latent nodes (see Fig.~\ref{fig:weakstrong:weak} and ~\ref{fig:weakstrong:strong}, respectively). The reason for proposing different models is that, as noted in~\cite{centeno2024significance}, when considering causal structures involving nonclassical latent nodes, the presence of intermediate latent nodes has an operational impact in terms of the set of achievable probability distributions.

\citet{minati2025experimental} pointed out that this variety of eavesdropping models is a novelty of networks, lacking analogue in the standard \textsf{Bell} scenario, albeit this discrepancy between the standard \textsf{Bell} scenario and networks only arises within the paradigm of having private settings. In randomness certification, we have emphasized that the measurement settings need not be private when assuming a ``listening'' adversary~\cite{pironio2013security}. Consequently, the possibility of considering an (overly) strong adversary \emph{does} arise in the standard \textsf{Bell} scenario upon restricting to a ``listening'' adversary. In the causal scenario, the intermediate latent nodes may have as parents both the latent root node and also some observed root node corresponding to a setting. 

\begin{figure}[h]
        \subfloat[\label{fig:strong_A}]{
                \centering
                \includegraphics[width=0.33\linewidth]{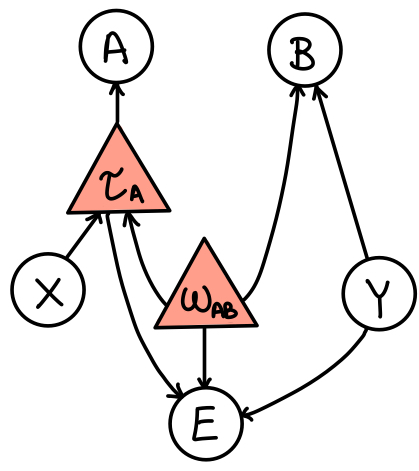}}
        \subfloat[\label{fig:strong_B}]{
                \includegraphics[width=0.33\linewidth]{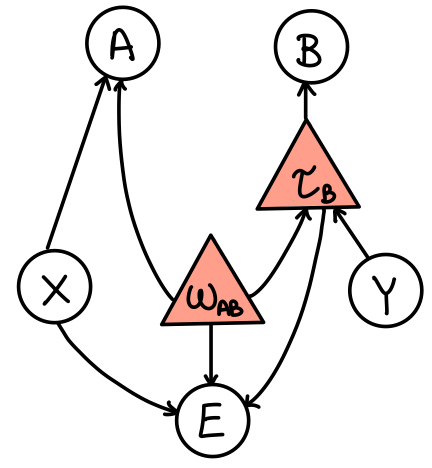}}
        \subfloat[\label{fig:strong_AB}]{
                \centering
                \includegraphics[width=0.33\linewidth]{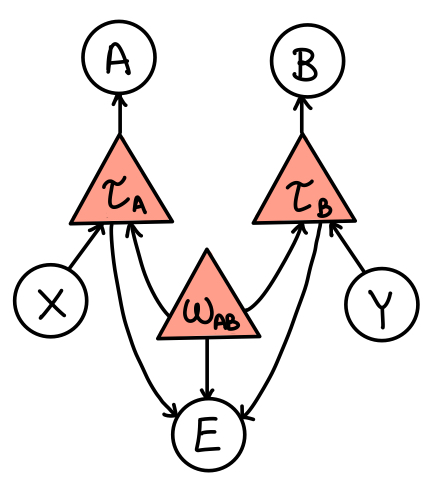}}
        \caption{Various ways of adding the presence of a strong eavesdropper on top of the standard \textsf{Bell} scenario within the paradigm wherein the eavesdropper \emph{may learn the values of the parties' settings}. In these variants of the standard \textsf{Bell} scenario no randomness can be certified for (a) Alice, (b) Bob, and (c) both. We endorse the public-settings paradigm, noting that the assumption of closure of laboratories fortunately then precludes us from needed to be concerned about strong eavesdropping.}
\label{fig:Bell-strong} 
\end{figure}

In Fig.~\ref{fig:Bell-strong}, we present the distinct possibilities of intermediate latent nodes for the standard \textsf{Bell} scenario with an eavesdropper who has access to the settings, i.e., the possible causal structures when considering a strong eavesdropper in the standard \textsf{Bell} scenario. However, even though randomness certification in the standard \textsf{Bell} scenario has been studied extensively, the possibility of a strong eavesdropper has never been considered -- and with good reason: In the standard \textsf{Bell} scenario, the strong eavesdropper model is so strong that it completely prevents \emph{any} randomness certification. Indeed, we will shortly argue that randomness certification is impossible in \emph{any} causal scenario, network or otherwise, when allowing for strong eavesdropping in the sense of Ref.~\cite{minati2025experimental}. Related to that consequence, we first argue that strong eavesdropping should be considered as violating the fundamental assumption of closure of laboratories.

Firstly, consider a scenario (for instance, \textsf{Bell}) without any eavesdropper. Look at any node (say, $A$) in that causal structure which corresponds to the outcome of a measurement. This observed node may have multiple parent nodes in the DAG, such as a setting, or one or more latent nodes corresponding to sources. Now, consider the following operation to create a different DAG:
\begin{samepage}
\begin{compactenum}
\item Add a new classical-type latent node $\lambda_A$ to the DAG, with an arrow $\lambda_A\to A$. 
\item For every parent node, $\text{node}_i$, of $A$ in the DAG, replace the arrow $\text{node}_i\to A$ with the arrow $\text{node}_i\to \lambda_A$.
\end{compactenum}
\end{samepage}
The resulting DAG now has $\lambda_A$ as a \emph{classical intermediate latent}. Per ~\cite{centeno2024significance}, this alternative DAG can explain exactly the same set of observable correlations as the original DAG; no more, and no less. This is an immediate consequence of the fact this this intermediate latent node has only one child, and hence removing it via the exogenization procedure would restore the original DAG while preserving observational equivalence.

Now, one way to reproduce the correlations in the original DAG using causal models in the modified DAG is to have $\lambda_A$ depend causally on its parents in the same manner as $A$ would do it in the original-DAG causal model. Then, in the modified-DAG causal model, $A$ depends deterministically on $\lambda_A$, namely, $A$ copies the value of $\lambda_A$. Such a causal model is really a trivial use of the intermediate latent; it is just a hidden copy of the outcome $A$ itself!

As such, it shows that allowing the eavesdropper to access such intermediate latents is the same as allowing the eavesdropper to listen in on the very outcomes of Alice's measurement themselves. But privacy of the measurement outcomes (albeit not of their causal antecedents) is the crux of the closure of laboratories assumption. Then, providing Eve with a copy of the measurement outcome, even if that copy is generated at a point in time before Alice observes and records her outcome, should also constitute a violation of the closure of laboratories assumption.

It should also be clear that with unlimited intermediate latents being available to Eve, we can always find a causal model in which the intermediate latents encode the outcomes of the later measurement deterministically, and hence the strong eavesdropping paradigm would prevent any randomness certification of any party in any network unless the settings are privileged as private, hence disallowing many intermediate latent varieties. Summarizing:

\begin{proposition}
     In the paradigm wherein the parties' settings are accessible to the eavesdropper, there is no randomness for \emph{any} party against a strong eavesdropper. Furthermore, even in the paradigm wherein settings are privileged relative to sources and treated as forever private, there is no randomness for any \emph{settingless} party against a strong eavesdropper.
\end{proposition}

\begin{figure}[h]
\centering
\includegraphics[width=5cm]{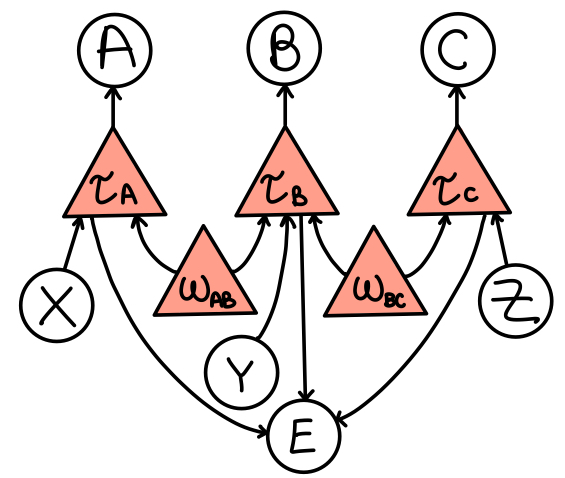}
\caption{Representation of the \textsf{bilocality} scenario relative to a strong eavesdropper, within the paradigm wherein the eavesdropper may learn the values of the parties' settings. Here there is no possibility of randomness. We reject this concern by appealing to the assumption of closure of laboratories, which precludes strong eavesdropping. Disallowing strong eavesdropping  salvages the possibility of randomness with the paradigm of settings being accessible to the eavesdropper.}
\label{fig:bilocality_all_intermediates}
\end{figure}

\begin{figure}[h]
\centering
\includegraphics[width=6cm]{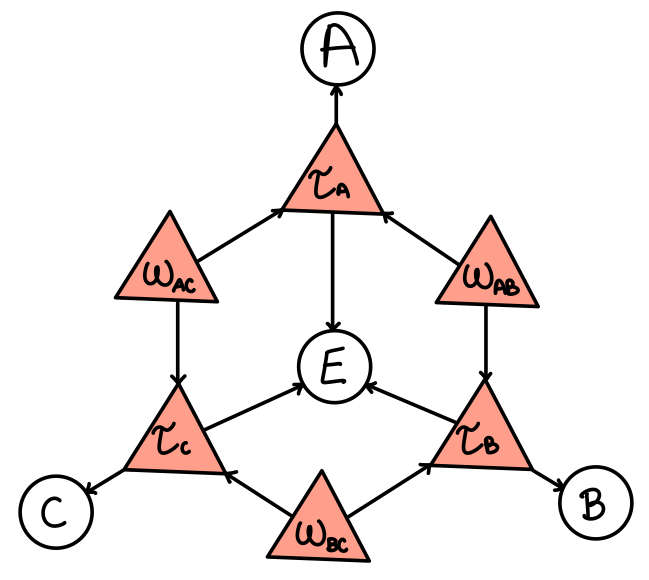}
\caption{Adding a strong eavesdropper to the \textsf{triangle} scenario is causally depicted in terms of intermediate latents. Here there is no possibility of randomness. We reject this concern by appealing to the assumption of closure of laboratories, which precludes strong eavesdropping. }
\label{fig:triangle-intermediates}
\end{figure}

Examples of causal structures with a strong eavesdropper within the paradigm where settings are eavesdropper accessible are given in Figs.~\ref{fig:bilocality_all_intermediates} and \ref{fig:triangle-intermediates} for the \textsf{bilocality} and the \textsf{triangle} scenario, respectively.

Thus we conclude that the strong eavesdropper should not be considered as a possible adversary model, as from our point of view, the minimal assumption of closure of laboratories is essential for device independent randomness certification.

That said, strong eavesdropping \emph{is} a sensible security concern within the paradigm (which we do not endorse) wherein settings are granted privileged private security status. That is, if even the strong eavesdropper \emph{never} may learn the values of the settings, then randomness remains plausible for any party with such a private setting. This is why~\citet{minati2025experimental} report nontrivial randomness relative to a strong eavesdropper in the \textsf{bilocality} scenario; their security analysis is depicted in Fig.~\ref{fig:weakstrong:strong}.

\section{Example of certifying randomness in the \textsf{bilocality} scenario using inflation technique}
\label{app: bilocality_inflation}
\begin{figure}[t]
\centering
\includegraphics[width=4.8cm]{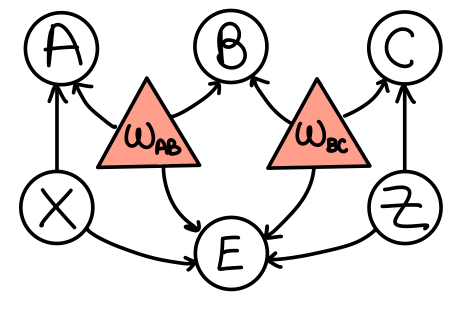}
\caption{Representation of the \textsf{bilocality} scenario with a listening eavesdropper, referred to as the \textsf{bilocality+E} scenario.}
\label{fig:bilocality-Eve}
\end{figure}

In this appendix, we explain in detail how the inflation technique works for the case of the \textsf{bilocality} scenario with an eavesdropper, $\textsf{bilocality+E}$ (Fig.~\ref{fig:bilocality-Eve}). For the sake of simplicity, let us consider the case where we want to certify single-party randomness using inflation. Then, in order to check if a particular probability distribution compatible with the \textsf{bilocality} scenario, $P^{obs}_{A,B,C|X,Z}$, exhibits intrinsic randomness in Alice we solve:
\begin{subequations}
 \begin{align}
        p_{\text{guess}}^{A_x}\coloneqq \; &\max\; \sum_a P_{A,E|X}(a,a|x)\\
        &\mbox{s. t.} \quad P_{A,B,C,E|X,Z} \in \mathcal{B}_E\\
        &\mbox{and} \quad P_{A,B,C|X,Z} = P_{A,B,C|X,Z}^{obs}\,,
    \end{align}
\end{subequations}

where $\mathcal{B}_E$ denotes the set of correlations that can be produced in the $\textsf{bilocality+E}$ causal structure (Fig.~\ref{fig:bilocality-Eve}). As explained before, we use the inflation technique to bound the set $\mathcal{B}_E$. Each level of the inflation hierarchy yields a tighter bound.

\begin{figure}[b]
\centering
\includegraphics[width=4.8cm]{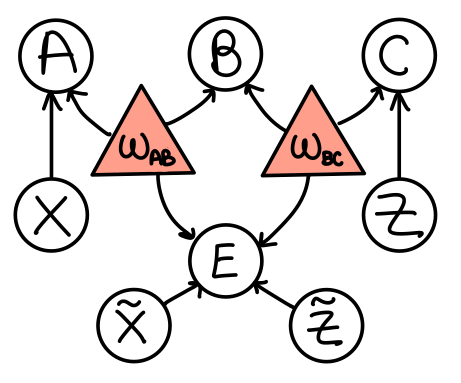}
\caption{The maximal interrupted DAG of the \textsf{bilocality+E} scenario. The interruption emphasizes that Eve can change the settings by which she measures the sources \emph{independently} of the settings of Alice and Bob.}
\label{fig:bilocality-interrupted}
\end{figure}

Let us now explain in detail some of the first levels of the inflation hierarchy. The first level of the inflation technique (i.e., when we consider scenarios that can be constructed using only one copy of the different devices) corresponds to considering the maximal interrupted DAG\footnote{This level is trivial in the cases where there is no variable to interrupt.} (Fig.~\ref{fig:bilocality-interrupted}). 

In order to pass this level of inflation (that is, the probability distribution $P \in \mathcal{B}_E^{(1,1)}$ where the superindex indicates the number of copies of each nonclassical source), one must be able to find a probability distribution $Q_{A,B,C,E|X,Z,\tilde{X},\tilde{Z}}$ such that it satisfies all the no-signaling constraints of the maximal interrupted DAG along with the compatibility constraints relative to the original scenario. Mathematically,

\begin{subequations}
\begin{align}
\nonumber
     &P_{A,B,C,E|X,Z} \in \mathcal{B}_E^{(1,1)}
     \mbox{iff} \\
     \nonumber \exists \;\;&Q_{A,B,C,E|X,Z,\tilde{X},\tilde{Z}}\quad \mbox{such that }\\
     &Q_{A,B,C|X,Z,\tilde{X},\tilde{Z}}=Q_{A,B,C|X,Z} \\
    &Q_{A,B,E|X,Z,\tilde{X},\tilde{Z}}=Q_{A,B,E|X,\tilde{X},\tilde{Z}}\\
    &Q_{B,C,E|X,Z,\tilde{X},\tilde{Z}}=Q_{B,C,E|Z,\tilde{X},\tilde{Z}} \\
    \begin{split}
    &Q_{A,B,C,E|X,Z,\tilde{X},\tilde{Z}}(a,b,c,e|x,z,x,z)\\
    &\quad \quad \quad = P_{A,B,C,E|X,Z}(a,b,c,e|x,z).
    \end{split}
\end{align}
\end{subequations}

Note that in practice, one solves the previous linear program when solving the optimization problem defined in Eq.~\eqref{optimization_networks} particularized for the \textsf{bilocality} scenario.

\begin{figure}[t]
\centering
\includegraphics[width=4cm]{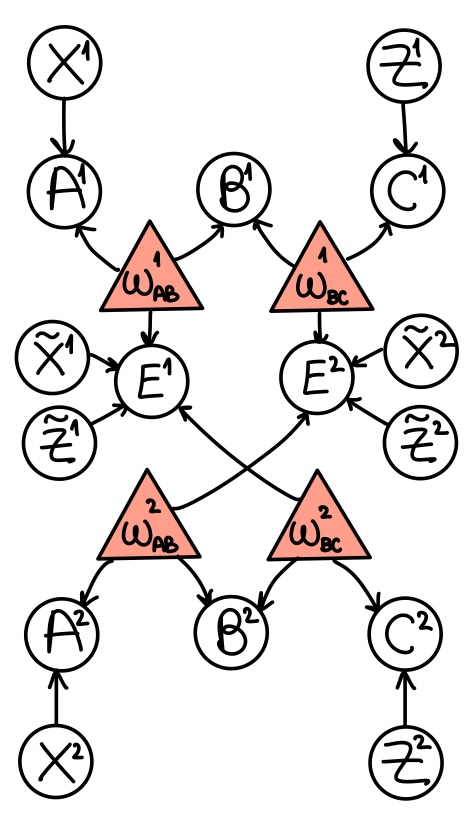}
\caption{Nonfanout inflation of the interrupted \textsf{bilocality+E} scenario.}
\label{fig:bilocality-inflations}
\end{figure}

Let us now consider the inflation level (2, 2), which means that two copies of each source are utilized to construct the inflated scenarios. That is, we define the set $\mathcal{B}_E^{(2,2)}$. As this is a higher inflation level, the previous one is included. This can be seen in the fact that we construct inflations which are maximally interrupted as well. For this particular example, at this level of inflation, there is only one nontrivial\footnote{Nontrivial inflations are those different from just a number of independent copies of the original network.} inflated network (Fig.~\ref{fig:bilocality-inflations}). Notice that this nontrivial inflation appears because of the eavesdropper, as the \textsf{bilocality} scenario (without any eavesdropper) does not have any nontrivial nonfanout inflation.

Then, following the same reasoning as in the first level, a probability distribution $ P_{A,B,C,E|X,Z}\in \mathcal{B}_E^{(2,2)}$ if there exists a probability distribution $Q$ over the observed nodes in the inflated scenarios that satisfies two sets of constraints: (i) the no-signaling conditions imposed by the inflated causal structure, ii) the compatibility constraints with respect to the original scenario. %
The no-signaling conditions are straightforward and therefore omitted for brevity. In this case, in contrast to the first level, the compatibility constraints are more complex, as they require that certain marginal distributions in the inflated scenario match those in the original one. Specifically, they apply to sets of variables whose associated subgraphs in the inflated DAG are structurally identical (i.e., isomorphic) to the original causal structure. These sets are referred to as \emph{injectable sets}\footnote{The formal definitions of injectable set and the concrete isomorphism used to say that two subnetworks are structurally identical  are given in~\cite{Wolfe_2019}.}. Moreover, these constraints can be subsumed in the constraints from the \emph{maximal injectable sets} - that is, the largest distinct sets of variables whose associated subgraphs in the inflated DAGs are structurally identical (i.e., isomorphic) to (a subgraph of) the original causal structure. Mathematically, for the inflation of Fig.~\ref{fig:bilocality-inflations} these are
\begin{align*}
    &Q_{A_iB_iC_i|X_iZ_i}(a,b,c|x,z) = P_{ABC|XZ}(a,b,c|x,z)\\
    &Q_{A_i E_i C_j|X_i Z_j\tilde{X}_i\Tilde{Z}_i}(a,e,c|x,z,x,z)=P_{AEC|XZ}(a,e,c|x,z),
\end{align*}
where $i\neq j$ and $i,j \in \{1,2\}$. %

In general, there could be more than one inflated scenario for a given network at a certain level. In those cases, one could consider any of the inflations individually to derive constraints on the set of feasible distributions. However, to fully exploit that level of inflation, all inflations must be considered simultaneously. Therefore, one has to add a third type of constraints: (iii) the cross-inflation constraints. They follow the same idea of matching marginals over structurally identical sets of nodes but, in this case, the isomorphism must be between the different inflations. Again, we can take the maximal isomorphic sets to subsume all these constraints (for this type of constraints, it is not needed to consider the injectable sets, as they are already taken into account in the compatibility constraints).

\begin{figure}[t]
\centering
\includegraphics[width=8.7cm]{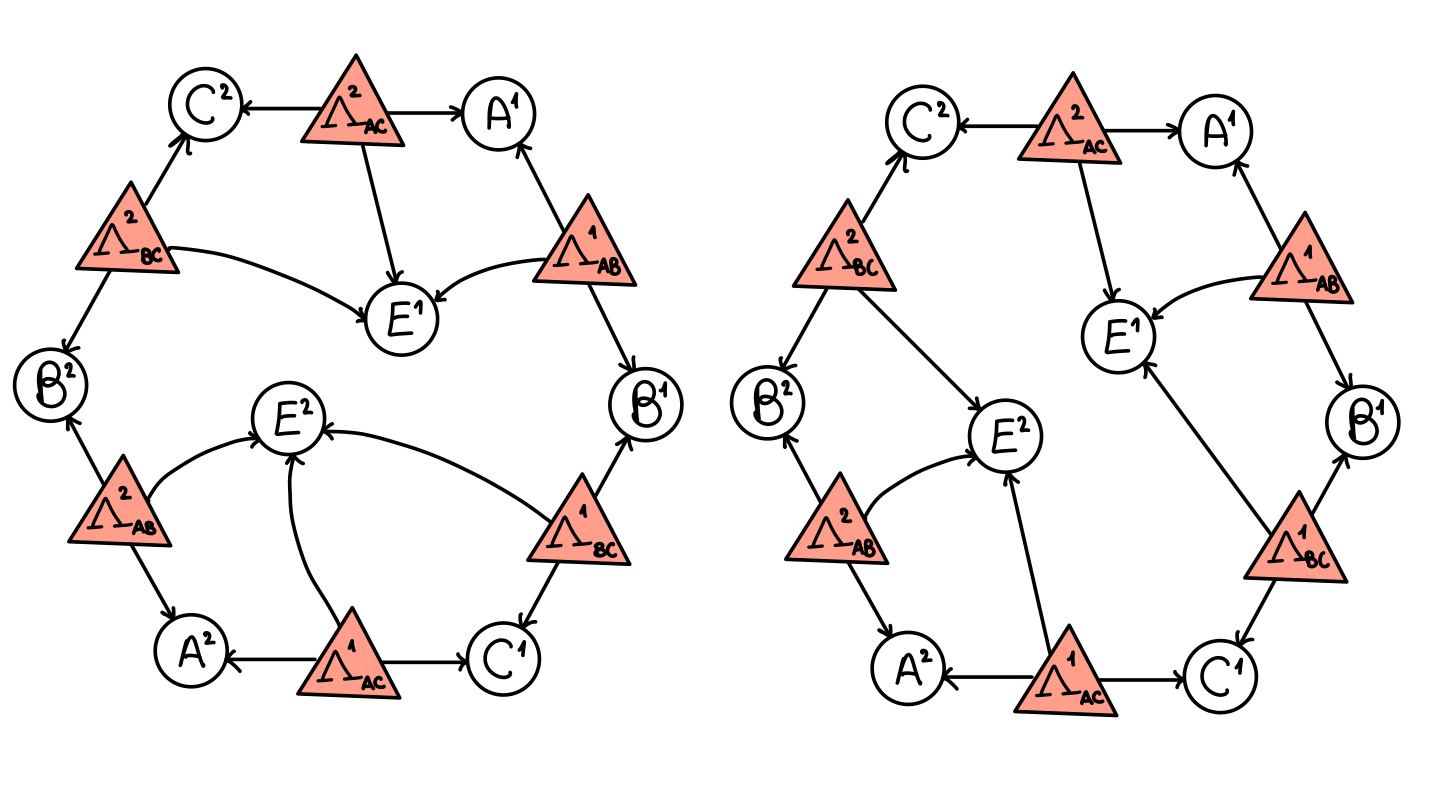}
\caption{Two of the possible nonfanout inflation of the \textsf{triangle+E} scenario.}
\label{fig:triangle-inflations}
\end{figure}

As an example, the second level of inflation of the \textsf{triangle+E} scenario has several inflations. Two of them are represented in Fig.~\ref{fig:triangle-inflations} and their corresponding cross-inflation constraint is:

\begin{align}
    Q^{12}_{A_1, B_1, C_1, A_2, B_2, C_2} = Q^{34}_{A_3, B_3, C_3, A_4, B_4, C_4},
\end{align}
where $Q^{12}$ and $Q^{34}$ are the probability distributions for the left-hand side and the right-hand side inflations of Fig.~\ref{fig:triangle-inflations}, respectively.

\section{Probability distributions}
\label{app: correlations}
This appendix provides all the probability distributions used in the paper for ease of reference. 
\subsection{Fritz's inspired correlation in the bilocality scenario}
This correlation is produced by a protocol inspired by~\cite{Fritz_2012}. In this protocol, Bob and Charlie share a classical source $\Lambda_{BC}$, which randomly sends the values 0 or 1. Bob's measurement is determined by $\Lambda_{BC}$ and Charlie ouputs $\Lambda_{BC}$ directly, ignoring $Z$. Then, Charlie's outputs can be interpreted as Bob's inputs. Also, Alice and Bob perform the measurements that violate CHSH: $\hat{A_0} = \sigma_Z$ and $\hat{A_1} = \sigma_X$, and $\hat{B_0} = (\sigma_Z+\sigma_X)/\sqrt{2}$ and $\hat{B_1} = (\sigma_Z-\sigma_X)/\sqrt{2}$. Thus, producing a nonclassical correlation in the \textsf{bilocality} scenario. Mathematically, this correlation can be written as follows:
\begin{align*}
    &P_{A,B,C|X,Z} = P_{C|Z}\cdot P_{A,B|X,C}\quad
    \mbox{where}\quad P_{C|Z} = \frac{1}{2}\\ &\mbox{and}\quad P_{A,B|X,C}(a,b|x,c) = \begin{dcases*} \frac{2+\sqrt{2}}{8}\quad \mbox{if}\quad a \oplus b = x \cdot c \\
    \frac{2-\sqrt{2}}{8} \quad\mbox{if}\quad a\oplus b \neq x \cdot y.
    \end{dcases*}
\end{align*}

\subsection{Entanglement-swapping}
Entanglement-swapping is a well-known phenomenon that generates nonclassicality in the \textsf{bilocality} scenario~\cite{branciard2010characterizing, branciard2012bilocal}. It involves establishing nonclassical correlations between two particles that have never interacted previously. For this protocol, the sources emit pairs of particles in a maximally entangled state, say $\ket{\phi^+} = (\ket{00}+\ket{11})/\sqrt{2}$. Bob performs a coarse-grained Bell state measurement on the two received particles, yielding two possible outputs $b=0,1$, which correspond to the measurement operators $\hat{B}_0 = \ket{\psi^+}\bra{\psi^+}$ and $\hat{B}_1 = \mathbb{1} - \ket{\psi^+}\bra{\psi^+}$, respectively. Then, when Bob outputs 0, he performs entanglement-swapping, and Alice and Charlie will be sharing a maximally entangled state. On the other hand, Alice and Charlie perform the measurements in a way that when Bob outputs 0, they can violate the CHSH inequality. In particular, the measurements are $\hat{A_0} = \sigma_Z$ and $\hat{A_1} = \sigma_X$, $\hat{C_0} = (\sigma_Z+\sigma_X)/\sqrt{2}$ and $\hat{C_1} = (\sigma_Z-\sigma_X)/\sqrt{2}$.

\subsection{Fritz's triangle correlation}
This correlation was proposed by Fritz in~\cite{Fritz_2012} and is a non-classical correlation that can be produced in the \textsf{triangle} without inputs (and four outputs for each party). In this protocol, two of the parties, let say Alice and Bob, violate the CHSH inequality using the sources that are not shared between them as inputs (that is, Alice uses $\Lambda_{AC}$ as input, while Bob uses $\Lambda_{BC}$). To do so, they share a maximally entangled state, say $\ket{\phi^+} ) (\ket{00}+\ket{11})/\sqrt{2}$, and perform the measurements that maximally violate CHSH ($\hat{A_0} = \sigma_Z$ and $\hat{A_1} = \sigma_X$, and $\hat{B_0} = (\sigma_Z+\sigma_X)/\sqrt{2}$ and $\hat{B_1} = (\sigma_Z-\sigma_X)/\sqrt{2}$). Both Alice and Bob will output the outcome of the measurement and the input used ($\Lambda_{AC}$ for Alice and $\Lambda_{BC}$ for Bob). Meanwhile, Charlie is used to certify the independence of the inputs that Alice and Bob use, therefore he outputs $\{\Lambda_{AC}, \Lambda_{BC}\}$.
\subsection{Post-quantum correlation}
This correlation is compatible with the settingless \textsf{triangle} with binary outputs and was proposed in~\cite{pozas2023post}. It is a nonclassical correlation that is not compatible with quantum theory and it is described as a network analogue of the Popescu-Rohrlich box. Mathematically, the probability distribution in terms of the correlations is as follows
\begin{equation*}
    P_{A,B,C}(a, b, c) = \frac{1}{8}\left[ 1+ (a+b+c)E_1 + (ab+ac+bc)E_2 + abcE_3 \right],
\end{equation*}
where $E_1 = 0 = E_3$ and $E_2 = \sqrt{2}-1$.

\section{Feasibility problems to certify lack of randomness despite requiring some nonclassical parents}
\label{app: lack_of_randomness_embeddings}

This appendix provides the feasibility problems to certify lack of randomness in the middle party for the bilocality scenario where Bob does not have settings and for the triangle scenario without settings. 

Formally, there can be no randomness for (settingless) Bob from $P_{A,B,C|X,Z}$ in the \textsf{bilocality} scenario if:

\begin{subequations}\label{LP_biloc_B}
\begin{align}\nonumber
\exists\;\; &\{Q_{A,B,C_0,C_1|X}\geq 0, \quad Q'_{A,B,E|X,Y,S}\geq 0\}\quad\textrm{ s.t. }\\
\nonumber &\eqref{LP_biloc_NSX}\text{, }\eqref{LP_biloc_fac}\text{, }\eqref{LP_biloc_rec}\text{,}\quad\text{and}\\
\begin{split}
\label{LP_biloc_B_rec}&Q'_{A,B|X,Y}(a,b|x,y{=}\{c_0 c_1\})\\
&\quad=Q_{A,B|X,C_0,C_1}(a,b|x,c_0,c_1)
\end{split}\\
\label{LP_biloc_B_predic}&Q'_{B,E|Y,S}(b,e|y,s{=}y)=P_{B}(b)\delta_{b,e}\\
\label{LP_biloc_B_NSX}&Q'_{B,E|X,Y,S}=Q'_{B,E|Y,S}\\
\label{LP_biloc_B_NSY}&Q'_{A,E|X,Y,S}=Q'_{A,E|X,S}\\
\label{LP_biloc_B_NSL}&Q'_{A,B|X,Y,S}=Q'_{A,B|X,Y}
\end{align}
\end{subequations}

Note that the extra constraints in Eqs.~\eqref{LP_biloc_B} do not constitute a distinct feasibility problem relative to that in Eqs.~\eqref{LP_biloc}; rather, the extra constraints in Eqs.~\eqref{LP_biloc_B} constitute a restriction on the space of feasible $Q_{A,B,C_0,C_1|X}$ beyond the minimal restrictions captured in Eqs.~\eqref{LP_biloc}.

\begin{figure}[b]
\centering
\includegraphics[width=3.5cm]{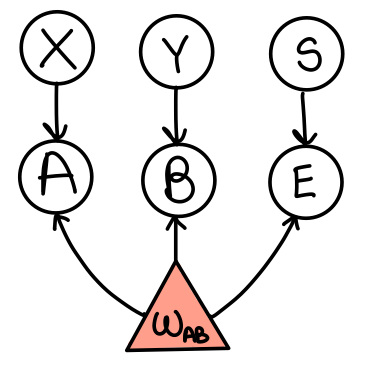}
\caption{Adding an eavesdropper to the \textsf{Bell} scenario while endowing Eve with her own setting. The setting for the eavesdropper is important to include when the setting for Bob is hidden, and yet we insist that Eve be able to perfectly predict Bob's outcomes by means of Bob's hidden setting and Alice's hidden setting being coordinated by a latent classical source.}
\label{fig:tripartite_Bell_Eve}
\end{figure}

Constraint \eqref{LP_biloc_B_rec} explicitly reinterprets the $Q_{A,B|X,C_0,C_1}$ correlation as a bipartite correlation where $Y$ is the hidden setting for Bob which is ultimately determined by classical latent source $\Lambda_{BC}$. Constraint~\eqref{LP_biloc_B_predic} enforces that the model should allow Eve to perfectly predict Bob's outcome. In that light, note that constraint~\eqref{LP_biloc_B_predic} only imposes agreement between Bob and Eve when their individual hidden settings coincide (as both hidden settings are determined by $\Lambda_{BC}$). Here, we have used the notation that $Y$ is the hidden setting of Bob and that $S$ is the hidden setting of Eve (see Fig.~\ref{fig:tripartite_Bell_Eve}). The cardinality of that hidden setting is the cardinality of $\Lambda_{BC}$, and we are implicitly presuming $|\Lambda_{BC}|=|C|^{|Z|}$ without loss of generality. 
Finally, constraints~\eqref{LP_biloc_B_NSX},~\eqref{LP_biloc_B_NSY}~and~\eqref{LP_biloc_B_NSL} enforces $Q'_{A,B,E|X,Y,S}$ to be nonsignalling.

We can also apply this idea to the \textsf{triangle} scenario. We can show that there is no randomness with respect to Bob in the triangle scenario if:

\begin{subequations}\label{LP_triangle_B}
\begin{align}\nonumber
\exists\;\; &\{ Q_{A,B,C_1,C_2,...,C_{d}|X}\geq 0,\; Q'_{Z}\geq 0,\; Q^{\prime\prime}_{A,B,E|X,Y,S}\geq 0\} \quad\textrm{ s.t.}\\ 
\nonumber &\eqref{triangle_NSX}\text{, }\eqref{triangle_fac}\text{, }\eqref{triangle_rec}\text{,}\quad\text{and}\\
\begin{split}
\label{LP_triangle_B_rec}&Q^{\prime\prime}_{A,B|X,Y}(a,b|x,y{=}\{c_1 c_2 ... c_d\})\\
&\quad=Q_{A,B|X,C_1,C_2,...,C_d}(a,b|x,c_1,c_2,...,c_d)
\end{split}\\
\label{triangle_B_predic}&Q^{\prime\prime}_{B,E|Y,S}(b,e|y,s{=}y)=P_{B}(b)\delta_{b,e}\\
\label{triangle_B_NSX}&Q^{\prime\prime}_{B,E|X,Y,S}=Q^{\prime\prime}_{B,E|Y,S}\\
\label{trianglec_B_NSY}&Q^{\prime\prime}_{A,E|X,Y,S}=Q^{\prime\prime}_{A,E|X,S}\\
\label{triangle_B_NSL}&Q^{\prime\prime}_{A,B|X,Y,S}=Q^{\prime\prime}_{A,B|X,Y}
\end{align}
\end{subequations}
As in Eqs.~\eqref{LP_biloc_B}, in Eqs.~\eqref{LP_triangle_B} we continue to employ the notation that $Y$ is the hidden setting of Bob and that $S$ is the hidden setting of Eve. Once again, the cardinality of that hidden setting is the cardinality of $\Lambda_{BC}$, and we are implicitly presuming $|\Lambda_{BC}|=|C|^{|Z|}$ without loss of generality. In contrast to Eqs.~\eqref{LP_biloc_B}, for the \textsf{triangle} scenario the cardinality $|Z|=d$ is an adjustable parameter of the model search space, as $Z$ is not actually observed in the \textsf{triangle} scenario.

It is worth emphasizing an important point. Since Charlie only receives classical information, both feasibility programs (Eqs.~\eqref{LP_biloc_B} and \eqref{LP_triangle_B}) can be adapted (by enlarging the cardinality of Eve) so that one can certify the absence of randomness for both Bob and Charlie simultaneously. Furthermore, one could also adapt the programs to certify the lack of randomness of the three parties simultaneously. The modification consists of requiring that Eve not only guesses Bob's outcome perfectly but also Alice's one in the embedded Bell scenario (again, this would require to increase the cardinality of Eve). In this way, as Charlie is only receiving classical information and we ensure that Alice and Bob are predictable by virtue of the embedding, we certify lack of randomness for the three of them.

\end{document}